\documentclass[apjl,twocolumn]{emulateapj}
\usepackage{natbib}
\newcommand{\lsim}{\raisebox{-0.13cm}{~\shortstack{$<$ \\[-0.07cm] $\sim$}}~}
\newcommand{\gsim}{\raisebox{-0.13cm}{~\shortstack{$>$ \\[-0.07cm] $\sim$}}~}

\newcommand{\tfm}{24 \, \rm  \mu m}

\shorttitle{A Generalized Power-Law Diagnostic for Infrared Galaxies at $\lowercase{z}>1$}
\shortauthors{Caputi}

\begin{document}

\title{A Generalized Power-Law Diagnostic for Infrared Galaxies at $\lowercase{z}>1$: \\ Active Galactic Nuclei and Hot Interstellar Dust}


\author{K. I. Caputi\altaffilmark{1}
}



\altaffiltext{1}{Kapteyn Astronomical Institute, University of Groningen, P.O. Box 800, 9700 AV Groningen,
The Netherlands. \\ Email: karina@astro.rug.nl}

\begin{abstract}

I present a generalized power-law diagnostic that allows to identify the presence of active galactic nuclei (AGN) in infrared (IR) galaxies at $z>1$, down to flux densities at which the extragalactic IR background is mostly resolved.  I derive this diagnostic 
from the analysis of 174 galaxies with $S_\nu(\tfm)> 80 \, \rm \mu Jy$ and spectroscopic redshifts $z_{\rm spec}>1$ in the Chandra Deep Field South, for which I study the rest-frame UV/optical/near-IR spectral energy distributions (SEDs), after subtracting a hot-dust, power-law component with three possible spectral indices $\alpha=1.3$, 2.0 and 3.0. I obtain that 35\% of these $\tfm$ sources are power-law composite galaxies (PLCGs), which I define as those galaxies for which the SED fitting with stellar templates, without any previous power-law subtraction, can be rejected with $> 2 \sigma$ confidence. Subtracting the power-law component from the PLCG SEDs produces stellar-mass correction factors $<1.5$ in $>80\%$ of cases. The PLCG incidence is especially high (47\%) at $1.0<z<1.5$. To unveil which PLCGs host AGN, I conduct a combined analysis of 4Ms X-ray data, galaxy morphologies, and a greybody modelling of the hot dust. I find that:  1) 77\% of all the X-ray AGN in my $\tfm$ sample at $1.0<z<1.5$  are recognised by the PLCG criterion; 2)  PLCGs with $\alpha=1.3$ or 2.0 have regular morphologies and $T_{\rm dust} \gsim 1000 \,\rm K$, indicating nuclear activity. Instead,  PLCGs with $\alpha=3.0$ are characterised by disturbed galaxy dynamics, and a hot interstellar medium can explain their dust temperatures $T_{\rm dust} \sim 700-800 \,\rm K$. Overall, my results indicate that the fraction of AGN among $\tfm$ sources is between $\sim$30\% and 52\% at $1.0<z<1.5$.

\end{abstract}

\keywords{infrared: galaxies  - galaxies: evolution - galaxies: high-redshift}

\section{Introduction}
\label{sec-intro}

The study of the individual dust-obscured sources that make the extragalactic infrared (IR) background \citep{pug96,dol06} has made enormous progress over the last decade, since the launch of the {\em Spitzer Space Telescope} \citep{wer04}, and successively  the {\em Akari Telescope} \citep{mur07} and the {\em Herschel Space Observatory} \citep{pil03}.  In particular, the scientific output of these missions has revealed the importance of powerful, dust-obscured star-formation and nuclear activity in shaping galaxy evolution at high redshifts ($z>1$). This activity was dominated by luminous and ultra-luminous infrared galaxies  \citep[LIRGs and ULIRGs;][]{lefl05,cap07},  which had a main role in the global star formation history of the Universe \citep[e.g.][]{hop06}, and the process of massive galaxy buildup at $z>1$ \citep{cap06a}.

The availability of multi-wavelength ancillary data from deep galaxy surveys has been crucial to investigate the presence and properties of IR galaxies at high $z$. Most redshift estimates and the derivation of other  parameters, such as the galaxy stellar mass, rely on the fitting of spectral energy distribution (SED) templates to broad-band photometry that traces the galaxy rest UV/optical and near-IR light ($\lambda_{\rm rest} \lsim 3 \, \rm \mu m$). The derivation of reliable values for these galaxy parameters requires a proper wavelength coverage of the photometric data, and also considering the possible variations that the galaxy SEDs may have, through the choice of sufficiently representative galaxy templates, and the analysis of departures of these SEDs from models of pure stellar evolution.

A long-standing problem regarding the composition of the extragalactic IR background is understanding to which extent dust-obscured active galactic nuclei (AGN) are part of the IR galaxy population at high redshifts.  A direct way to reveal nuclear activity is searching for X-ray detections \citep[e.g.][]{rig04,pol06,fio08,sym11,mat12}, but dust-obscured AGN can remain undetected even in typically deep X-ray maps. Spectral line diagnostics at optical and IR wavelengths are also useful to recognise AGN, but they are usually limited to relatively small samples of bright IR sources \citep[e.g.][]{arm07,saj07,cap08,des08,nar08,her09,pet11,hwa12}.

An alternative method that has proven to be quite efficient to reveal AGN in large samples of dust-obscured galaxies is the use of mid-IR colour-colour diagrams \citep{lac04,ste05}, based on photometry at $3.6-8.0 \, \rm \mu m$, taken with the {\em Spitzer} Infrared Array Camera \citep[IRAC;][]{faz04}. The segregated locus that some AGN occupy in these diagrams is related to the fact that pure AGN SEDs are characterised by a power-law shape at rest-frame optical/near-IR wavelengths. In fact, the presence of a pure IRAC power-law SED has been proposed by some authors as a criterion to select AGN-dominated galaxies among {\em Spitzer} $\tfm$ and IRAC-selected sources \citep{alo06,don07}. But AGN selection through IRAC colour-colour diagrams has well-known limitations: it is complete only for bright IR sources \citep[e.g.][]{lac07}, and the resulting samples are typically contaminated by star-forming galaxies \citep{don12}. This is especially the case at high redshifts, as the galaxy stellar emission is shifted into the IRAC bands, producing similar IRAC colours to those of AGN at lower redshifts.

In addition to the identification, there is a second problem, which consists in understanding what fraction of the galaxy light  is due to the AGN component at different wavelengths.  At mid- and far-IR wavelengths ($3 \lsim \lambda_{\rm rest} \lsim 1000 \, \rm \mu m$), disentangling the star-formation/AGN contributions is required to properly derive the on-going, obscured star formation rates in IR galaxies. This issue has been tackled in different studies of LIRGs and ULIRGs at high $z$, by analysing mid-IR spectroscopic data, or mid-/far-IR broad-band data \citep[e.g.][]{fad10,gru10,bar12,han12,mul12,poz12}.

In the rest-frame optical/near-IR ($0.3 \lsim \lambda_{\rm rest} \lsim 3 \, \rm \mu m$), a power-law component associated with an AGN  distorts the light of the underlying stellar populations in the host galaxy, and may affect the derived galaxy stellar mass. For galaxies that are identified with X-ray luminous AGN, the SED fitting with stellar templates, and derivation of stellar masses are usually avoided \citep[e.g.][]{cap06b}, or attempted only after subtracting empirical AGN templates \citep[e.g.][]{mer10,mai11}. In all other cases, however, the derivation of stellar masses is of common practice, even when there is the suspicion that the rest-frame near-IR light could be partly contaminated by an AGN component (for example, in the case of an IRAC-band excess in the galaxy SED). Quantifying the importance of this effect in IR-selected galaxies is then necessary to understand the reliability of the derived host galaxy properties.

In this work I present the analysis of the rest-frame optical/near-IR SEDs of 174 $\tfm$-selected galaxies with $S_\nu(24 \, \rm \mu m)\geq 80 \, \rm \mu Jy$  and secure spectroscopic redshifts $z_{\rm spec}>1$. My aim is to investigate whether, and when, a power-law component makes a relevant contribution to the galaxy SED, i.e. the subtraction of a power-law component produces a significant improvement of the galaxy SED fitting with stellar templates.  Hereafter, I will refer to these galaxies as {\em power-law-component galaxies} (PLCGs). Note that this criterion is more relaxed than the IRAC power-law shape imposed by other authors to select potential IR AGN candidates \citep{alo06}, and therefore, the sample analysed here should also include galaxies for which the AGN emission is less dominant over the underlying host galaxy light.

This paper is organised as follows.  In Section \S\ref{sec-sample}, I give details on the sample selection and compilation of  multi-wavelength photometry. In Section \S\ref{sec-seds}, I present the results of the SED analysis, and quantify the effect on the derived stellar masses.  Later, in Section \S\ref{sec-nature}, I investigate the nature of the PLCGs at $z>1$. I make use of  ultra-deep X-ray data to obtain an independent diagnostic of nuclear activity among the $\tfm$ sources, and understand in which cases the significant SED power-law component is related to the presence of an AGN. I also analyse the PLCG morphologies, and derive the greybody temperatures that are necessary to explain the SED power law. In Section \S\ref{sec-iraccol}, I investigate how the latest proposed IRAC colour-colour criteria deal with the selection of PLCGs and other X-ray-detected AGN. Finally, in Section \S\ref{sec-summ}, I summarise my findings and present some concluding remarks.  All magnitudes and colours quoted in this paper are total and refer to the AB system \citep{oke83}.  I adopt a cosmology with  $H_0=70 \,{\rm km \, s^{-1} Mpc^{-1}}$, $\rm \Omega_M=0.3$ and $\rm \Omega_\Lambda=0.7$. All stellar masses refer to a Salpeter (1955) initial mass function (IMF) over star masses of $(0.1-100) \, \rm M_\odot$.

\section{Sample selection and multi-wavelength datasets}
\label{sec-sample}

The Great Observatories Origins Deep Survey (GOODS) programme \citep{gia04} comprises a wide range of deep galaxy surveys conducted with main astronomical facilities, including the {\em Spitzer} and {\em Hubble Space Telescopes}, the {\em Chandra X-ray Observatory}, and the largest optical/near-IR ground-based telescopes. In particular, the {\em Spitzer} images for the GOODS fields have been collected as part of the GOODS {\em Spitzer} Legacy Programme (PI: M. Dickinson), and include deep data from both IRAC and the Multiband Imaging Photometer for {\em Spitzer} \citep[MIPS;][]{rie04}. 

The 174 $\tfm$-selected galaxies analysed here have been extracted from the publicly released  $\tfm$ catalogue of the GOODS-South (GOODS-S) field, which has a flux density limit of $S_\nu(24 \, \rm \mu m)=80 \, \rm \mu Jy$, down to which this catalogue is highly reliable and complete\footnote{see http://irsa.ipac.caltech.edu/data/GOODS/docs/goods\_dr3.html}. Sources down to this flux density limit make $\sim 70\%$ of the extragalactic $\tfm$ background \citep{pap04,dol06}.

To decide which $\tfm$ sources would be part of the analysis sample, I applied the following two criteria: 1) the source should have a secure (good-quality flag)  spectroscopic redshift ($z_{\rm spec}$) from any of the multiple spectroscopic galaxy surveys of the GOODS-S field (Le F\`evre et al.~2004; Szokoly et al.~2004; Teplitz et al.~2007; Vanzella et al.~2008; Popesso et al.~2009; Fadda et al.~2010; Kurk et al.~2013); 2) the redshift should be $z_{\rm spec}>1$. I searched for spectroscopic counterparts of the $\tfm$ sources within a 1.5 arcsec matching radius, and considered only one-to-one identifications. The final sample contains 174 $\tfm$ sources with $z_{\rm spec}>1$, with similar numbers of galaxies at $1.0<z<1.5$ and $z \geq 1.5$ (89 and 85 sources, respectively).

Restricting the SED study only to galaxies with secure spectroscopic redshifts ensures that all the results and conclusions in this paper are free from the uncertainties that are usually introduced by photometric redshift determinations at high redshifts. On the other hand, the multiple GOODS-S spectroscopic datasets correspond to a wide variety of selection criteria, and even include some IR spectra \citep{fad10}, so the sub-sample of $\tfm$ galaxies with spectroscopic redshifts is reasonably representative of the entire sample of $S_\nu(24 \, \rm \mu m)>80 \, \rm \mu Jy$ galaxies at $z>1$. This has been verified by comparing the  $S_\nu (24 \, \rm \mu m) / S_\nu(i \, \rm band)$ colour distributions of the current spectroscopic sample and the overall GOODS-S $\tfm$ sample with $z>1$, with spectroscopic and photometric redshifts (Caputi et al.~2006a,b; 2007). 

\begin{figure}
\plotone{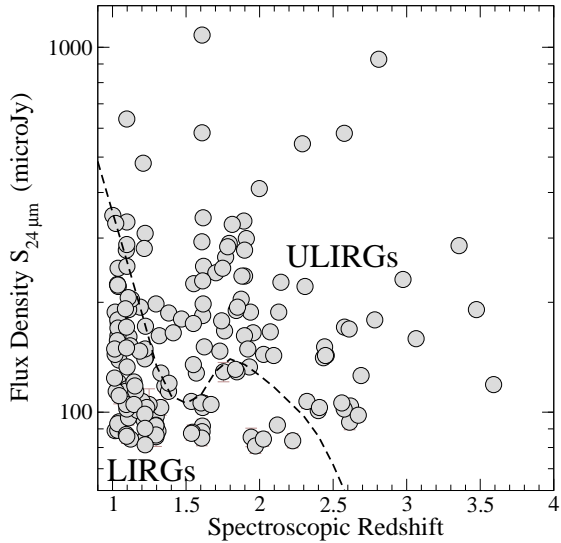}
\caption{Flux densities $S_\nu(24 \, \rm \mu m)$ vs. spectroscopic redshifts $z_{\rm spec}$ for the 174 galaxies in the $\tfm$ sample with $z_{\rm spec}>1$. The dashed line, which divides the LIRG and ULIRG regimes,  has been computed by considering the Bavouzet et al.~(2008)  monochromatic-to-total IR luminosity conversion, re-calibrated  for galaxies with $\nu L_\nu^{\rm 8 \, \mu m}> 10^{10} \, \rm L_\odot$ \citep[see][]{cap07}.  The k-correction factors are based on a mixture of IR star-forming galaxy templates \citep{lag04} and empirical IR spectra. \label{fig_stfmvsz}}
\end{figure}

The $S_\nu(24 \, \rm \mu m)$ flux densities versus spectroscopic redshifts $z_{\rm spec}$ of the 174 galaxies are shown in Fig.~\ref{fig_stfmvsz}. The dashed curve in this plot separates the LIRG and ULIRG regimes at bolometric IR luminosities $L_{IR}=10^{12} \, \rm L_\odot$. To compute total IR luminosities, I have considered  the Bavouzet et al.~(2008)  monochromatic-to-total IR luminosity conversion, re-calibrated  for galaxies with $\nu L_\nu^{\rm 8 \, \mu m}> 10^{10} \, \rm L_\odot$ \citep[see][]{cap07}. As in this work, the k-correction factors are based on a mixture of IR star-forming galaxy templates \citep{lag04} and empirical IR spectra. Although, in a strict sense, this LIRG/ULIRG separation curve corresponds to star-forming galaxies, it is suitable for the purpose of illustrating the appoximate regions of the LIRG and ULIRG regimes in the $S_\nu(24 \, \rm \mu m)$-$z$ diagram. 
 
\vspace{0.2cm}

Independently, I made use of the publicly released GOODS-S IRAC maps and the SEXTRACTOR software \citep{ba96} to extract a catalogue of IRAC sources at 3.6 and $4.5 \, \rm \mu m$. I measured the photometry at 5.8 and $8.0 \, \rm \mu m$  using SEXTRACTOR in dual-map mode. In order to construct a multi-wavelength optical/near-IR catalogue for the IRAC-selected sources, I searched for counterparts of these sources in the GOODS-S {\em Very Large Telescope (VLT)} ISAAC near-IR images, and the {\em Hubble Space Telescope} Advanced Camera for Surveys (ACS) maps, within a matching  radius of 0.5 arcsec. The finally compiled catalogue includes photometry in 11 bands: $B$, $V$, $i$, $z$, $J$, $H$, $K_s$, [3.6], [4.5], [5.8] and [8.0].  All the magnitudes in this catalogue are total and have been corrected for galactic extinction. To obtain total magnitudes, I considered aperture magnitudes in circles of 4-arcsec (IRAC) and 2-arcsec (ISAAC and ACS) diameter, and applied the corresponding aperture corrections.

Within the catalogue of IRAC-selected sources with multi-wavelength photometry, I identified all the 174 $\tfm$ galaxies  with spectroscopic redshifts $z_{\rm spec}>1$. The SED fitting analysis explained in next section is based on this photometry. A total of 22 galaxies (12.5\% of the sample) are out of the field of view in all the ACS bands, or at least one of the ISAAC bands. For this minority of sources, the SED fitting is based on 7,8,9 or 10 bands, depending on the case. For all the remaining galaxies, the SED fitting is based on the 11-band photometry. In case of a non-detection, a 2$\sigma$ upper limit has been considered for the flux density in the corresponding band.

\section{Spectral Energy Distribution Analysis}
\label{sec-seds}

\subsection{The Incidence of a Power-Law Component in the SEDs of Infrared Galaxies}
\label{sec-powlaw}

I considered that the flux density $S_{\nu} (\lambda)$ of each galaxy at observed wavelengths $\lambda \leq 8 \, \rm \mu m$ can be decomposed as:

\begin{equation}
S_{\nu} (\lambda) = S_{\nu}^{\rm stell.} (\lambda) +  b S_{\nu} (8 \, \rm \mu m) \times \left(\frac{\lambda}{8 \, \rm \mu m}\right)^\alpha, 
\end{equation}

\noindent where the two terms on the right-hand side correspond to a stellar component and a power-law component with spectral index $\alpha$, respectively. The power-law term is normalised to the $8 \, \rm \mu m$ flux density, so the constant $b$ can vary between 0 and 1.  This decomposition is similar to that proposed by Hainline et al.~(2011) to analyse the rest optical/near-IR SEDs of sub-millimetre galaxies at $z\sim2$.

\begin{figure}
\epsscale{.90}
\plotone{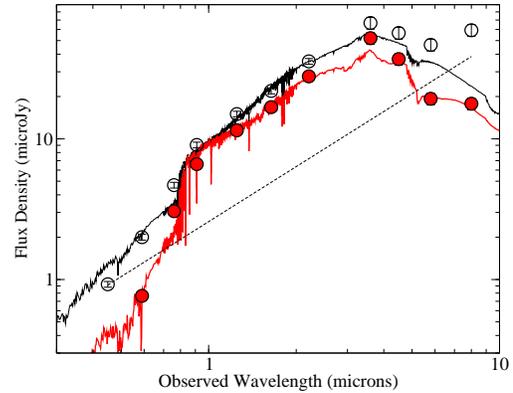}
\caption{Illustration of the technique applied here to identify galaxies with a significant SED power-law component. A power-law (dashed line) is subtracted from the original photometry of a galaxy (open circles), producing a new photometric dataset (filled circles). Stellar templates are then fitted to the original and power-law subtracted photometry (thin black and thick red lines, respectively). For each galaxy, 31 power-law subtractions with different combinations of $\alpha=1.3, 2.0$ and $3.0$, and $b=0.0,0.1, 0.2, \dots ,1.0$,  have been tested. In the example shown here, it can be seen that a power-law subtraction improves the overall SED fitting, particularly allowing for a better agreement with the datapoints in the IRAC bands. \label{fig_exam}}
\end{figure}

For each of the 174 galaxies in my $\tfm$ sample, I  produced a suite of 31 photometric sets $S_{\nu}^{\rm stell.} (\lambda)$ in the $BVizJHK_s$[3.6][4.5][5.8][8.0] bands, corresponding to different power-law subtractions. I considered that $\alpha$ could take three, and $b$ ten possible values: $\alpha=1.3, 2.0$ and $3.0$,  and $b=0.0, 0.1, 0.2, \dots , 1.0$. The case with $b=0.0$ and any $\alpha$ value corresponds to the original photometry $S_{\nu}^{\rm stell.} (\lambda) = S_{\nu} (\lambda)$.  The three adopted $\alpha$ values are representative of the IR power-law indices that characterise different types of AGN, with different hot-dust components (e.g.  Alonso-Herrero et al.~2003, 2006; Polletta et al.~2006; H\"onig et al.~2010).

I performed the SED modelling on the 31 photometric sets $S_{\nu}^{\rm stell.} (\lambda)$ of each galaxy, using a customised $\chi^2$-minimisation SED-fitting code that incorporates the 2007 version of the Bruzual \& Charlot synthetic stellar template library, with solar metallicity \citep{bch03,bru07}. These templates correspond to different star formation histories: a single stellar population, and different exponentially-declining star-formation histories with $\tau=0.1$ through 5~Gyr. In all cases, the redshift of the galaxy has been fixed to the known $z_{\rm spec}$ value.  To account for internal extinction, I convolved the stellar templates with the Calzetti et al.~(2000) reddening law, allowing for $0.0\leq A_V \leq 3.0$ with a step of 0.1. Stellar masses are obtained in the output of the same SED-fitting code.

The technique is illustrated in Figure~\ref{fig_exam}: once a power-law component is subtracted from the original photometry (open circles), a new photometric dataset is obtained (filled circles). For each galaxy, I have tested 31 possible power-law subtractions (including the original photometry), as explained above. I have then fitted stellar templates to all these photometric sets, and compared the resulting minimum $\chi^2$ values.

\begin{figure}
\plotone{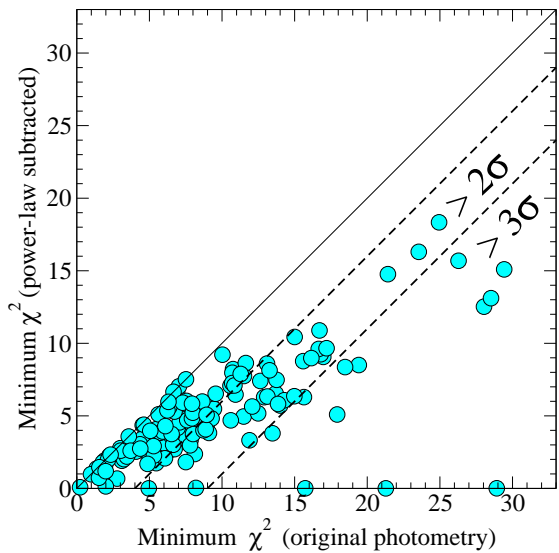}
\caption{Minimum $\chi^2$ value obtained in the SED fitting of $\tfm$ galaxies with $z_{\rm spec}>1$ using stellar templates, after subtracting a power-law component to the photometry, vs. the original value obtained with no power-law subtraction. The dashed lines delimit the regions with $\Delta \chi^2 > 4.0$ and 9.0, i. e. where the SED fitting with no previous power-law subtraction can be rejected with 2$\sigma$ and $3\sigma$ confidence, respectively. \label{fig_chi}}
\end{figure}

Figure~\ref{fig_chi} shows the minimum reduced $\chi^2$ value after power-law subtraction (i.e. the absolute minimum obtained considering all the different ($\alpha$, $b$)  combinations), versus the minimum reduced $\chi^2$ value obtained with the original photometry, for each galaxy. The points lying on the identity line correspond to galaxies for which the original photometry, without a power-law subtraction, produces the absolutely best fitting. The points below the identity line indicate the galaxies for which the SED fitting with stellar templates improves after the subtraction of a  power-law component. This is the case for the majority of the $\tfm$ galaxies. In particular, in 60 out of 174 cases (35\%), the best SED-fitting solution on the original photometry, without a power-law subtraction, can be rejected with $> 2\sigma$ confidence. These 60 galaxies are, according to my definition, the PLCGs in the $\tfm$ sample with $z_{\rm spec}>1$.

\begin{figure}
\plotone{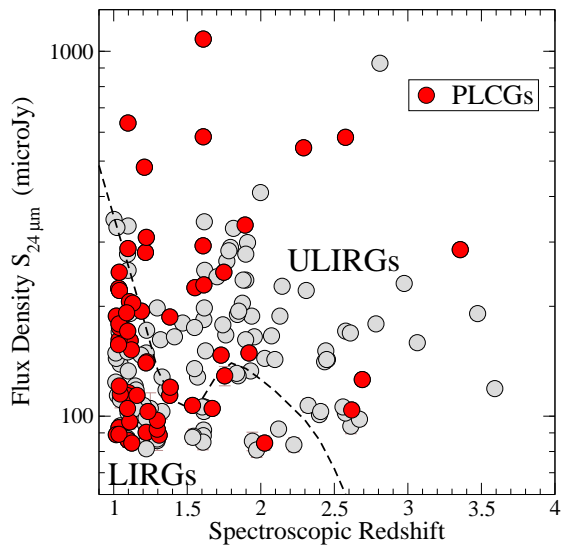}
\caption{Same as Fig.~\ref{fig_stfmvsz}, with the PLCGs highlighted. \label{fig_plcgstfmz}}
\end{figure}

Interestingly, around two-thirds of the PLCGs (42 out of 60) lie at $z_{\rm spec}<1.5$ (Fig.~\ref{fig_plcgstfmz}). This implies that $\sim 47\%$ of all the LIRGs and ULIRGs at $1.0<z<1.5$ are PLCGs, i.e. they have a significant power-law component in their rest-frame near-IR SEDs. The percentage of PLCGs at $1.0<z_{\rm spec}<1.5$ is much higher than the known percentage of AGN among LIRGs and ULIRGs at these redshifts, and raises the question of whether the power-law component is actually related to an AGN presence in all these cases. I investigate this issue further in Section \S\ref{sec-nature}. At $z>1.5$, instead, the fraction of PLCGs among $\tfm$ galaxies (the vast majority of which are ULIRGs) is of only $\sim 21\%$.

I repeated the same SED modelling using the 2003 version of the Bruzual \& Charlot stellar template library, and found very similar results:  $\sim 38\%$ of the $\tfm$ sources with $z_{\rm spec}>1$ are classified as PLCG. The resulting minimum $\chi^2$ values are very similar to those obtained with the 2007 templates (the median of the minimum $\chi^2$ differences is 0.009). All the following analysis will be based on the 2007 template library run, but no conclusion in this paper would change if I adopted the results obtained with the 2003 templates.

The host galaxy properties, as determined from the best-fitting stellar templates, are similar for PLCGs and non-PLCGs. For both sub-samples, the best-fitting star formation histories correspond to single stellar populations or $\tau$ models with $\tau \leq 0.1 \, \rm Gyr$ in $>80\%$ of cases. After the power-law subtraction, these models provide the best fitting for 93\%
 of the PLCGs. The best-fitting extinction values for PLCGs (before power-law subtraction) and non-PLCGs are also very similar: the median and r.m.s. of the distributions are $A_V=1.4\pm0.6$ and $A_V=1.4\pm0.7$, respectively. After power-law subtraction, the PLCGs have $A_V=1.0\pm0.6$.

\vspace{0.2cm}
\begin{figure}
\epsscale{.90}
\plotone{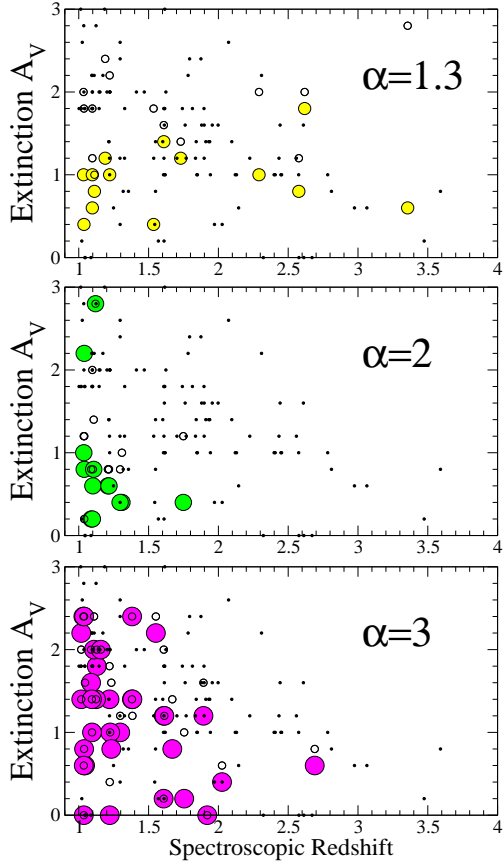}
\caption{Best-fitting extinction $A_V$  vs. spectroscopic redshifts for the 60 PLCGs, classified according to their spectral index $\alpha$. The $A_V$ values obtained before and after power-law subtraction are shown in each case (small open circles and larger coloured circles, respectively). The $A_V-z_{\rm spec}$ values of the non-PLCGs are shown as a reference in all panels (dots). \label{fig_zdisalpha}}
\end{figure} 

Figure \ref{fig_zdisalpha} shows the extinction $A_V$ versus spectroscopic redshifts for the 60 PLCGs, classified according to their power-law indices $\alpha$. The non-PLCGs are also shown as a reference. A bit more than a half of the PLCGs (55\%) are characterised by a power-law component with $\alpha=3$, while the remaining PLCGs have lower best-fitting power-law indices, i.e. $\alpha=1.3$ or 2. These indices are related to the temperature of the hot-dust component in the galaxy, which is higher for lower $\alpha$ values (cf. Section \S\ref{sec-gbtemp}).  

The effect of a decrease in the best-fitting $A_V$ value is especially evident for the $\alpha=1.3$ PLCGs: a shallow power-law component can mimic the effect of additional reddening. After power-law subtraction, the  majority of the $\alpha=1.3$ and $2$ PLCGs have $A_V \leq 1$, while most of the $\alpha=3$ PLCGs have $A_V>1$.

The normalisation factor of the power-law component is $b \geq 0.50$ for virtually all the PLCGs, which means that at least 50\% of the $8 \, \rm \mu m$ flux density is in the power-law component. For around a half of the PLCGs, this contribution is $\geq 80\%$. This confirms that the PLCG definition adopted here, based on the $> 2\sigma$ improvement of the SED fitting, truly selects sources for which the power-law component makes an important contribution to the rest-frame near-IR light of the galaxy.

\vspace{0.2cm}
\begin{figure}
\plotone{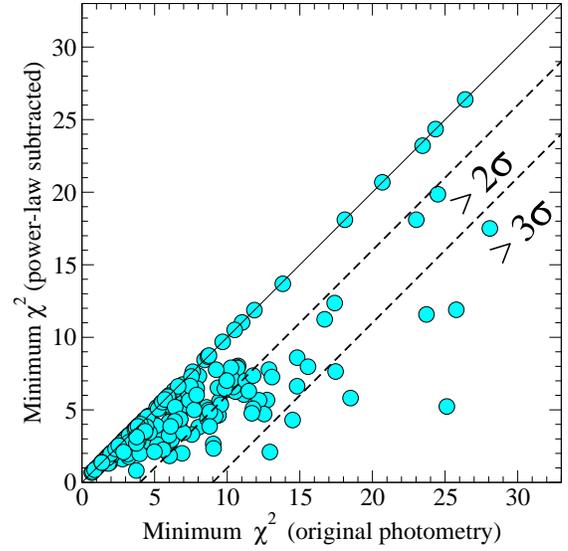}
\caption{Same plot as in Fig.~\ref{fig_chi}, but for the control sample of 247 IRAC galaxies with $z_{\rm spec}>1$ which are not $\tfm$-detected. \label{fig_chicon}}
\end{figure}

For galaxies at $z\sim1.4$, the emission from polycyclic aromatic hydrocarbons (PAHs) at rest $\lambda_{\rm rest}\sim 3.3 \, \rm \mu m$ will enter the observed $8 \, \rm \mu m$ band. Although this is a relatively faint PAH emission feature, it can potentially mimic the effect of a steep power-law component in the galaxy SED. To assess the importance of this effect on the PLCG selection, I have performed the SED fitting on additional photometric sets for each galaxy, with power-law subtractions corresponding to larger spectral indices, namely, $\alpha=3.5$ and $4$. Within the total sample, only 13 galaxies appear to be PLCGs with such large best-fitting spectral index. However, their redshifts are not concentrated around $z\sim1.4$ or any other specific redshift, so the steeper $\alpha$ values are unlikely the effect of emission features. In fact, 11 out of these 13 galaxies have been recognised as PLCGs with $\alpha=3$ in my original analysis. Only two galaxies appear as new PLCGs: one at $z_{\rm spec}=2.810$ (corresponding to an X-ray luminous AGN), and another one at $z_{\rm spec}=1.550$ (for which the $3.3 \, \rm \mu m$ PAH emission could be partly within the $8 \, \rm \mu m$-filter wavelength coverage). These results indicate that: a) the $3.3 \, \rm \mu m$ PAH emission plays a minor role in the overall IR SED for most IR galaxies; b) considering $\alpha>3$ values for the power-law subtraction produces a virtually negligible effect on the identified PLCG sample.

To investigate whether the high fraction of PLCGs is truly a characteristic of the $\tfm$ galaxy sample, I performed a similar SED fitting analysis on a control sample of 247 galaxies selected from the IRAC catalogue described in Section \S\ref{sec-sample}, also with secure $z_{\rm spec}>1$, but which are not $\tfm$-detected (i.e. $S_\nu(24 \, \rm \mu m) < 80 \, \rm \mu Jy$). These galaxies have been selected to have a similar stellar mass distribution as the $\tfm$ galaxies ($M_{\rm stell.} \gsim 10^{10} \, \rm M_\odot$, based on the SED modelling of the original photometry).  For each of these galaxies, I have produced 31 sets of photometric variations in the same manner explained above, and fitted the resulting SEDs with stellar templates in all cases. The comparison of minimum $\chi^2$ values with and without power-law subtraction is shown in Fig.~\ref{fig_chicon}.

\begin{figure*}
\epsscale{1.1} 
\plotone{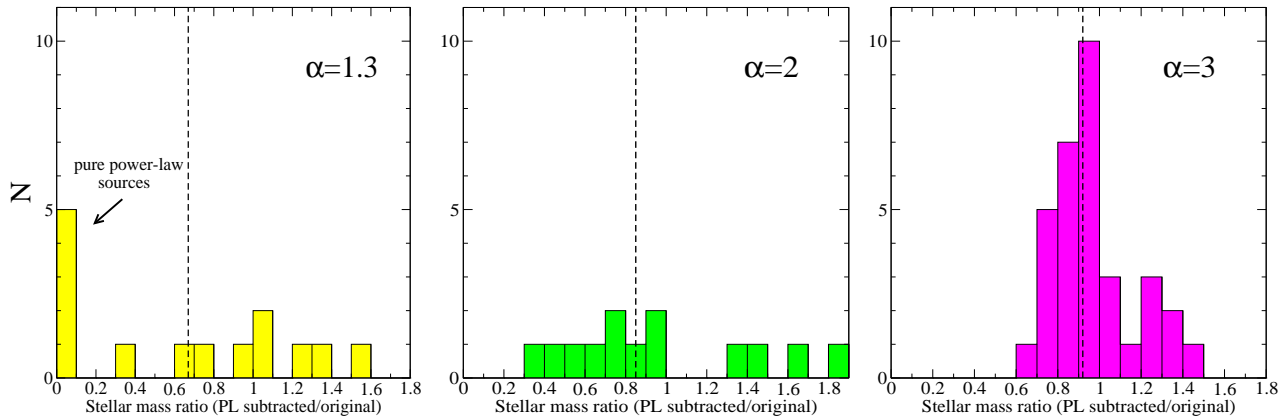}
\caption{Distribution of corrected-to-original stellar mass ratios for the PLCGs, classified according to their best-fitting power-law index $\alpha$. The corrected stellar masses are those derived from the SED modelling after the best-fitting power-law subtraction. The vertical, dashed lines indicate the median correction factors derived from each distribution. \label{fig_stmratio}}
\end{figure*}

For the control sample, the percentage of galaxies that significantly improve their SED fitting with stellar templates after a power-law subtraction is a factor of two smaller than for the $\tfm$ galaxies: a pure stellar SED fitting can be discarded with $> 2\sigma$ confidence in only 18\% of cases (44 out of 247 galaxies). These 44 PLCGs in the control sample have best-fitting $\alpha$ values shared in similar proportions as for the $\tfm$-detected PLCGs: 50\% have best-fitting $\alpha=3$, while the other 50\% have $\alpha=1.3$ or 2.  

These results obtained on the control sample indicate that the incidence of PLCGs among mid-IR-selected galaxies is twice more important than among other similarly massive galaxies at $z>1$.

\subsection{The Effect of a Power-Law Subtraction on the Derived Stellar Masses}
\label{sec-stm}

Figure \ref{fig_stmratio} shows the distribution of corrected-to-original stellar mass ratios for the 60 $\tfm$-selected PLCGs, classified according to their best-fitting power-law index. The corrected stellar masses are those derived from the SED modelling with stellar templates after the best-fitting power-law subtraction. 

The inspection of these distributions shows that the correction on the stellar masses is non-trivial: the exact correction needs to be derived on a case-by-case basis. However, for $>$80\% (90\%) of the PLCGs the correction is within a factor of $\sim 1.5$ (2.0).  The widest correction-factor distributions are those corresponding to power-law spectral indices $\alpha=1.3$ and 2, for which the stellar mass ratios range between $\lsim 0.1$ and $\sim 1.8$. The median values of these stellar mass ratios are 0.67 and 0.85, respectively.

The corrections on the stellar masses are the smallest for the case $\alpha=3$ (the median of the ratio distribution is 0.92). This behaviour is expected, as the subtraction of such a steep power-law has a rapidly declining effect on the photometry at wavelengths shorter than $8 \, \rm \mu m$, i.e the overall SED shape is not substantially modified except for a single photometric point.

It may a priori seem surprising that the SED fitting for some galaxies yields {\em larger} stellar masses after the PL subtraction. This effect is produced because the best-fitting stellar template properties (star formation history, age and extinction) are usually different for the PL-subtracted photometry to those for the original photometry, and, in some cases, the resulting overall normalisation factors (i.e. mass-to-light ratios) are larger.

For 5 out of the 60 PLCGs, all with best-fitting $\alpha=1.3$, the corrected stellar mass is less than $1/10$ of the stellar mass derived from the original photometry. These cases correspond to ``pure power-law sources'',  and all except one of them are identified as X-ray luminous AGN with $L_X > 10^{43} \, \rm erg \,s^{-1}$ in the Chandra 4~Ms catalogue for the GOODS-S field \citep{xue11}. The only source that is not X-ray detected is the highest-redshift of these pure power-law sources, with $S_\nu(24 \, \rm \mu m)\approx 286 \, \rm \mu Jy$ and $z_{\rm spec}=3.356$, so it is likely an intrinsically luminous AGN as well.  For these sources, a stellar mass calculation based on the original photometry should clearly be avoided. But, as said above, in all other cases the correction to the stellar mass is within a factor of two. This result validates the approach of obtaining approximate stellar mass estimates for e.g. low-luminosity AGN, using the original un-corrected photometry, that is followed by some authors \citep[e.g.][]{sil09,ros13}.

\section{The Nature of the PLCGs}
\label{sec-nature}

\subsection{Independent AGN Diagnostics: X-ray Detections}
\label{sec-xrays}

In order to investigate whether the PLCGs actually host AGN, I searched for counterparts of these sources in the publicly available, ultra-deep Chandra 4Ms X-ray catalogue for the GOODS-S field \citep{xue11}.  This diagnostic is very useful to assess to which extent the PLCG criterion can be used to identify AGN in the general case that such deep X-ray catalogues are not available (to date, X-ray maps of such a depth have only been obtained for the GOODS-S field).

\begin{figure}
\plotone{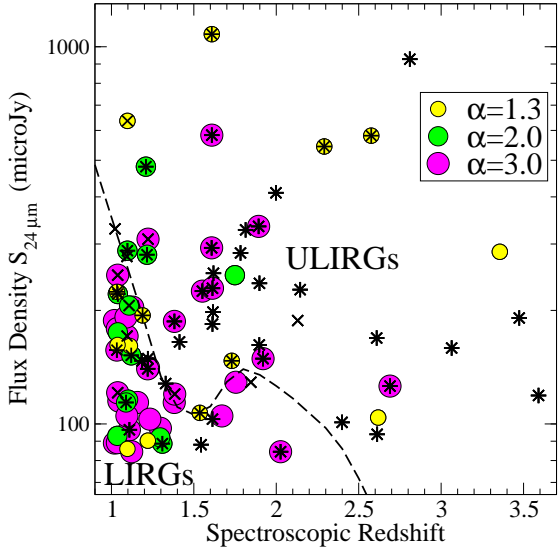}
\caption{Flux densities $S_\nu(24 \, \rm \mu m)$ vs. spectroscopic redshifts $z_{\rm spec}$ for the $\tfm$ galaxies at $z_{\rm spec}>1$ that are identified with X-ray sources in the Chandra 4Ms catalogue. Asterisks and crosses correspond, respectively, to AGN and normal galaxies, classified according to their X-ray properties. All PLCGs are shown with couloured circles.  \label{fig_xrayid}}
\end{figure}

Fig. \ref{fig_xrayid} shows the $S_\nu(24 \, \rm \mu m)$-$z$ diagram  for the $\tfm$ galaxies in the $z_{\rm spec}>1$ sample that are identified with X-ray sources in the Chandra 4Ms catalogue. The PLCGs are highlighted with couloured circles. There are a total of 59 $\tfm$ galaxies identified with an X-ray source, 48 out of which correspond to X-ray-classified AGN, and 11 are X-ray `normal galaxies'. The X-ray classification is based on the intrinsic X-ray luminosities, X-ray colours (hardness ratios), and derived neutral gas column densities \citep[e.g.][]{bau04,xue11}. Although the X-ray-based AGN classification is among the most secure, the criteria typically adopted to classify  X-ray sources could still result in some low-luminosity or highly-obscured AGN being classified as normal galaxies (see Xue et al.~2011 and references therein).

\vspace{0.2cm}

The analysis of Fig. \ref{fig_xrayid} shows the following.

\noindent $\bullet$ At $z<1.5$, 77\% of the X-ray AGN are recognised as PLCGs, which indicates that the PLCG criterion produces a high completeness in the identification of AGN at these redshifts. And, interestingly, among the X-ray-classified `normal galaxies', the ratio of PLCGs is very similar ($\sim 78\%$).  So, either these galaxies also host an AGN, or there is another mechanism that gives rise to the X-ray emission, and is also responsible for the hot-dust component manifested in the PLCG classification.  

In starburst galaxies, supernova explosions can produce hot-gas bubbles with significant X-ray emission. The resulting shock waves can heat the surrounding dust grains up to sublimation temperatures $T \gsim 1000 \, \rm K$. However, if  the X-ray luminosities of X-ray galaxies were only a product of supernova explosions, they would correlate with the IR luminosities, as both trace the on-going star formation rates \citep{mas08}. {\em This is not the case for the PLCGs at $z<1.5$, for which two sources can differ in more than a factor of three in IR luminosity, but both be classified as X-ray normal galaxies with similar X-ray luminosities.}  This becomes clear by inspection of Fig. \ref{fig_plcgstfmz}, where some bright $\tfm$ sources at $1.0<z<1.5$ are neither PLCGs nor X-ray galaxies at all.  Thus,  the SED power-law excess of these PLCGs is not simply tracing high star formation rates. 

At $z<1.5$, it is interesting to note the different distributions of best-fitting spectral indices $\alpha$ among the X-ray AGN and normal galaxies that are recognised as PLCGs: 9 out of 13  (69\%) X-ray AGN / PLCGs have $\alpha=1.3$ or 2, while only 2 out of 7 (29\%) of the X-ray galaxy / PLCGs have $\alpha=1.3$ or 2. Or, equivalently, only 31\% of the X-ray AGN classified as PLCGs have $\alpha=3$, while 71\% of the X-ray galaxies classified as PLCGs have $\alpha=3$. I will carry on this discussion in Section \S\ref{sec-morph}, where I will argue that the spectral index $\alpha$ is a good discriminator of PLCGs of different nature.
 
\vspace{0.2cm}

\noindent $\bullet$ Only 42\% of the X-ray-detected AGN at $z>1.5$ are classified as PLCGs. This effect could be due to a real difference in the properties of X-ray AGN at different redshifts: e.g. at $z>1.5$ the light of the host galaxies may dominate the rest near-IR SED, making a power-law component to appear less significant. Alternatively, the lower percentage of AGN incidence could be the consequence of a subtle k-correction effect, as the observed $8 \, \rm \mu m$ photometry only traces rest wavelengths $\lambda_{\rm rest} \lsim 3 \, \rm \mu m$  at $z>1.5$, while it traces up to rest $\lambda_{\rm rest} \sim 4 \, \rm \mu m$ at $z=1$. 

To investigate these two possible explanations, I consider those $\tfm$ galaxies at $z\sim2$ that form part of the Fadda et al.~(2010) IR galaxy sample. Pozzi et al.~(2012) have studied the mid-/far-IR SED decomposition of these galaxies, and found that nine out of 24 of them have a significant AGN component. Only three out of these nine sources are recognised as PLCGs within my sample. This result suggests that data beyond observed $8 \, \rm \mu m$ are necessary to reveal more effectively an AGN presence in galaxies at $z\sim2$.

In Section \S\ref{sec-gbtemp}, I will show that very hot dust temperatures ($T_{\rm dust} \gsim 1300 \, \rm K$) are necessary to generate a power-law SED component detectable in the IRAC bands at $z>1.5$. Such hot dust is typically found in AGN inner dusty tori \citep{bar87,mor12}, but some AGN are characterised by lower dust temperatures, or the inner torus may be less exposed \citep[e.g.][]{sch05}.  This will restrict the fraction of AGN that can be identified with my PLCG criterion, which is based on SED fitting up to observed $8 \, \rm \mu m$, at $z\sim2$.

\noindent $\bullet$ Around a half of the PLCGs at $1.0<z<1.5$ are not X-ray detected. The mean optical/near-IR SED properties of these PLCGs are not significantly different to those of X-ray-detected PLCGs classified as X-ray galaxies: e.g. the median extinction is $A_V=1.0$ for the X-ray undetected, against $A_V=0.8$ for the X-ray detected. Their IRAC colours are also similar (see Section \S\ref{sec-iraccol}).  The PLCG classification, and the X-ray detection, do not follow a simple correlation with the host galaxy IR luminosities: there are X-ray detected PLCGs both in the LIRG and ULIRG regimes. But all the X-ray undetected PLCGs are LIRGs, which suggests that the X-ray non-detection is due to the sensitivity limit of the X-ray surveys.

\subsection{PLCGs Non-Detected in X-Rays: Stacking Analysis}
\label{sec-plnonx}

To probe whether the X-ray non-detection of half of the PLCGs at $1.0<z<1.5$ is simply due to the sensitivity limit of the Chandra 4Ms X-ray survey, I performed a stacking analysis of these sources in X rays.

\begin{figure}[h]
\plotone{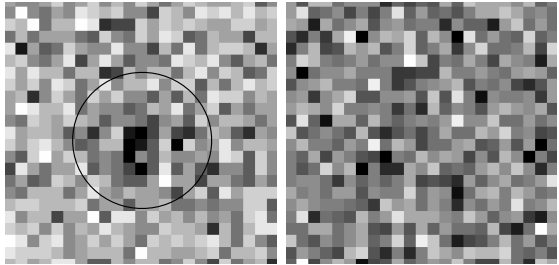}
\caption{Stacked X-ray images of the $1.0<z_{\rm spec}<1.5$  PLCGs that are not individually detected in X rays. {\em Left:} 0.5-2.0 keV; {\em right:} 2.0-8.0 keV. Each stamp covers $11 \times 11$~arcsec$^2$. \label{fig_xraystack}}
\end{figure}

\begin{figure*}
\epsscale{1.1} 
\plotone{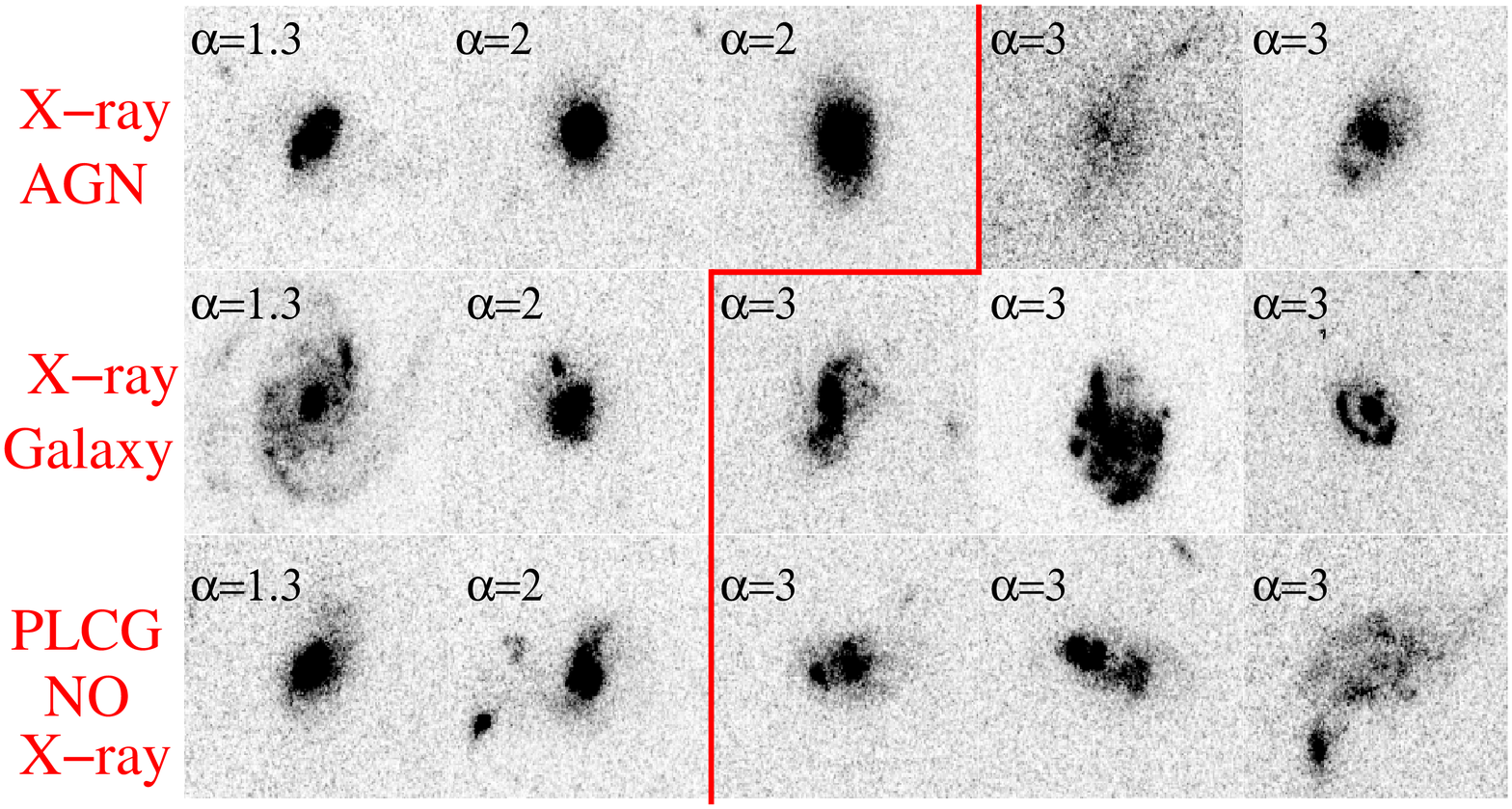}
\caption{{\em HST}/ACS $z$-band postage stamps of 15 PLCGs at $1.0<z<1.5$. Each stamp shows an area of $5 \times 5 \, \rm arcsec^2$ centred at the source.  The best-fitting spectral index $\alpha$ is indicated in each case. The solid lines separate the cases with $\alpha=3.0$, which have a much more disturbed morphology than the PLCGs with smaller spectral indices $\alpha=1.3$ and 2.0. \label{fig_stamps}}
\end{figure*}

Fig.~\ref{fig_xraystack} shows the stacked X-ray images of the PLCGs that are not individually detected in X rays at $1.0<z<1.5$. In the hard X-ray band (2.0-8.0 keV), the stacking does not produce any detection, while in the soft band (0.5-2.0 keV) a detection with a $> 6 \sigma$ significance is obtained. This stacked signal confirms that these sources  are X-ray emitters, but their fluxes are below the sensitivity limits of the 4Ms Chandra survey.

Nuclear activity cannot be discarded as an agent partly responsible for the soft X-ray emission in these sources, but it is probably not the main cause:  on average they have $\log_{10} (f_X/f_R) \approx -2.7$,  too low a value for an AGN classification \citep{xue11}.

Around two thirds of the stacked PLCGs have spectral index $\alpha=3$ which, as I discuss below, should mainly correspond to star-forming galaxies with hot interstellar media (ISM) at $1.0<z<1.5$. The stacking of the $\alpha=3$  PLCGs alone yields a $4.8\sigma$ detection in the (0.5-2.0 keV)  band, which corresponds to an average flux density $f_X \approx 1.9 \times 10^{-17} \, \rm erg \, cm^{-2} \, s^{-1}$ (after correction for galactic extinction; see Stark et al.~1992). At the median redshift of these sources $z_{\rm med}=1.13$, this corresponds to a luminosity $L_X \approx 1.3 \times 10^{41} \, \rm erg \, s^{-1}$, assuming a photon index $\Gamma=2$, which is appropriate for star-forming galaxies \citep[e.g.][]{leh08}. Using the relation calibrated by Mineo et al.~(2013), I find that this corresponds to an instantaneous star formation rate $\rm SFR \approx (37\pm4) \, \rm M_\odot \, yr^{-1}$. This is significantly lower than the SFR derived from the average total IR luminosities of these sources, which is $SFR \approx 90 \, \rm M_\odot \, yr^{-1}$, as obtained using the SFR-$L_{\rm IR}$ relation given by Kennicutt~(1998). This difference indicates the presence of internal absorption. To reconcile both SFR values, I find that these galaxies should have an average column density $N_H \approx 4.5 \times 10^{21} \, \rm cm^{-2}$.

In any case, as commented above, one has to be careful by considering the PLCG X-ray luminosity 
as a tracer of the instantaneous star formation rate, as many of these galaxies are likely composite star-forming/AGN systems. Also, the X-ray emission may be the result of energetic processes in the galaxy ISM that are not simply related to the on-going star formation rate. Some of these effects probably account for the dispersion observed in the $L_X$-SFR relation.

\subsection{Morphologies of PLCGs and X-Ray Galaxies at $1.0<z<1.5$}
\label{sec-morph}

Further clues about the nature of PLCGs, particularly at $1.0<z<1.5$, can be obtained by comparing their morphologies.   Figure \ref{fig_stamps} shows {\em HST}/ACS $z$-band postage stamps of 15 PLCGs at $1.0<z<1.5$, grouped in three different categories: X-ray detected sources, classified as AGN and Galaxies, and PLCGs that are not individually X-ray detected. The three classes display objects with different morphologies, including galaxies with compact and regular matter concentrations, and galaxies with very disturbed morphologies. The latter are seemingly the product of galaxy mergers, or are characterised by other violent processes, such as the radial ejection of material  (see the X-ray galaxy at the right end of the middle row).

The striking conclusion that can be extracted from Fig.\ref{fig_stamps} is the trend between the PLCG morphology and the value of the best-fitting spectral index $\alpha$: those PLCGs with $\alpha=1.3$ or 2.0 show quite a regular morphology, while the PLCGs with $\alpha=3.0$ are irregular and show signs of different kinds of galaxy interactions. 

This trend is independent of the X-ray detection and classification. This fact suggests that the spectral index $\alpha$ contains information about the nature of the source: in the PLCGs with $\alpha=1.3$ or 2.0, the SED power-law component is very likely the signature of nuclear activity, while in the $\alpha=3$ PLCGs the power law appears to be the consequence of a hot ISM produced by the violent galaxy dynamics. Shocks of hot gas can compress and heat up the ISM dust to several hundred Kelvin, and even up to sublimation  temperatures \citep[$T \gsim 1000 \, \rm K$;][]{ost06}. This hot dust, not related to an AGN origin, will also manifest itself as a power-law component in the rest near-IR galaxy SED. 

Of course, in some cases, an AGN and a host galaxy with a disturbed ISM may co-exist, producing X-ray confirmed AGN that are PLCGs with $\alpha=3$ (upper right stamps in Fig.~\ref{fig_stamps}; see also Kocevski et al.~2012 for a discussion on AGN morphology).

\begin{figure}
\plotone{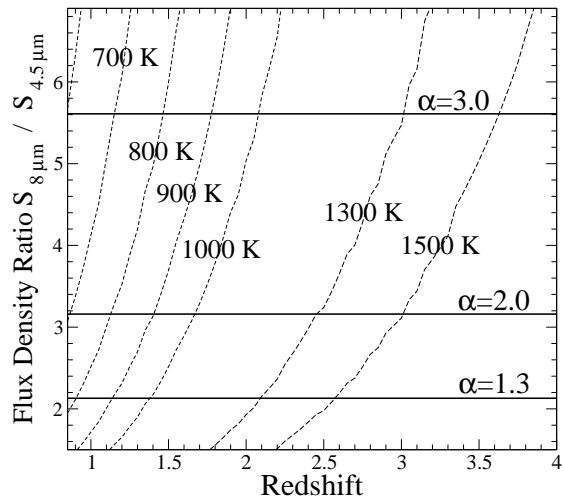}
\caption{Observed flux density ratios $S_\nu(8.0 \, \rm \mu m) / S_\nu(4.5 \, \rm \mu m)$  for greybody emission at different temperatures, as a function of redshift (dashed lines). The horizontal, solid lines indicate the fixed, observed ratios produced by power-law components with different spectral indices $\alpha$. \label{fig_coltemp}}
\end{figure}

\subsection{Estimated Temperatures of the PLCG Hot-Dust SED Components}
\label{sec-gbtemp}

I estimated the dust temperatures necessary to produce a power-law component in the galaxy SEDs, assuming a greybody function for the dust emission:

\begin{equation}
\label{eq-grey}
S_\nu \propto \lambda^{-\beta} \times \frac{2 h c}{\lambda^3} \, \frac{1}{e^{hc /\lambda k T} -1},
\end{equation}

\noindent where the second factor on the right-hand side of equation (\ref{eq-grey}) corresponds to the black-body emission at temperature $T$, and $h$ and $k$ are the Planck's and Boltzmann's constants, respectively. I assumed $\beta=1.5$, as it is common in the literature.

Fig. \ref{fig_coltemp} shows the observed flux density ratios $S_\nu(8 \, \rm \mu m)/ S_\nu(4.5 \, \rm \mu m)$ produced by greybody emitters at different temperatures, as a function of redshift. The horizontal solid lines indicate the fixed, observed flux density ratios produced by power-laws with the three spectral indices considered in this paper, i.e. $\alpha=1.3$, 2.0 and 3.0.  

Two main results can be derived from this simple diagram. Firstly, to generate a power-law component with $\alpha=1.3$ or 2 at $1.0<z<1.5$, temperatures as high as $900-1000 \, \rm K$ are needed. Instead, for a power-law with spectral index $\alpha=3.0$, temperatures $T \sim 700-800 \, \rm K$ are sufficient. This may explain why most of the X-ray AGN, for which dust temperatures close to the sublimation point are expected,  are associated with PLCGs with spectral indices $\alpha=1.3$ or 2, at $1.0<z<1.5$. The PLCGs with spectral index  $\alpha=3.0$ can be explained with a shock-heated ISM, with no need of dust getting close to the sublimation point (the exact dust sublimation temperature depends on the dust composition, but it is typically $T \gsim 1000 \, \rm K$). The existence of hot interstellar dust in IR galaxies has been previously discussed in the literature. Particularly, Lu et al.~(2003) found that a colour dust temperature of $\sim 750 \, \rm K$ can account for the near-IR excess observed in some local IR normal galaxies, similarly to what I find here for LIRGs at $1.0<z<1.5$.

Secondly, at redshifts $z>1.5-2.0$, producing a power-law component in the observed IRAC SED requires hot dust with temperatures $T \gsim 1300 \, \rm K$, for any of the spectral indices. This mostly explains why the incidence of PLCGs is lower at $z>1.5$, as only sources with such an extremely hot dust component will be classified as PLCG with the criterion adopted in this paper.

\section{IRAC Colour-Colour Diagnostic}
\label{sec-iraccol}

The use of {\em Spitzer}/IRAC colour-colour diagrams has proven to be quite successful for identifying AGN among IR-selected galaxies \citep{lac04,ste05}.  Over the years, these colour selections have been refined to be more reliable for AGN identification among fainter IR sources \citep[e.g.][]{don12}.  Here I analyse how the PLCGs, and other $\tfm$ galaxies at $z_{\rm spec}>1$, are classified in an IRAC colour-colour diagram. 

\begin{figure}
\plotone{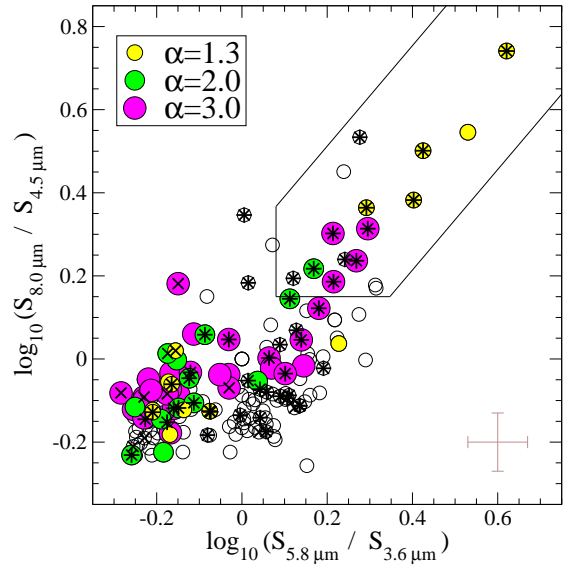}
\caption{IRAC colour-colour diagram for all the $\tfm$ sources with $z_{\rm spec}>1$ (circles). As in previous figures, PLCGs are highlighted in colour. Asterisks and crosses indicate, respectively, X-ray-classified AGN and galaxies. The solid lines delimit the colour region for AGN selection proposed by Donley et al.~(2012). Typical error bars on the colours are shown in the bottom right corner of the plot. \label{fig_iraccol}}
\end{figure}

Fig.~\ref{fig_iraccol} shows the IRAC colour-colour diagram for all the $\tfm$ galaxies at $z_{\rm spec}>1$, with the PLCGs and X-ray sources highlighted. The colours shown here are overall colours, i.e. there is no separation of stellar and power-law SED components. One can see that the range of overall IRAC colours displayed is quite wide, with some $\tfm$ galaxies having much redder IRAC colours than others.  PLCGs, in particular, display a wider range of colours than the non-PLCGs, as there are very few non-PLCGs with very red IRAC colours $\log_{10}(S_{8.0}/S_{4.5}) \gsim 0.2$ and $\log_{10}(S_{5.8}/S_{3.6}) \gsim 0.1$.  For PLCGs, I find no correlation between the IRAC colours and the spectral index of the SED power-law component. 

The solid lines in Fig.~\ref{fig_iraccol} delimit the revised colour region for AGN selection proposed by Donley et al.~(2012). More than 80\% of the $\tfm$ galaxies that lie within that region are PLCGs or X-ray AGN, indicating that indeed the proposed colour selection is highly reliable. However, these colour criteria are too strict and miss a substantial fraction of PLCGs, many of which are X-ray confirmed AGN. The incompleteness of this colour selection was discussed by Donley et al.~(2012), who argued that their criteria quickly lose completeness for AGN with X-ray luminosities $L_X<10^{44} \, \rm erg \, s^{-1}$. Within the current sample, all the X-ray-detected sources in the Donley et al.~(2012) region have $L_X \gsim 5 \times 10^{43} \, \rm erg \, s^{-1}$.

There are three galaxies that lie within the Donley et al. AGN colour-colour region, but which are neither identified as PLCGs, nor X-ray detected. One of them is a $z_{\rm spec}=1.045$ galaxy for which the SED fitting with a power-law subtraction just lies below the 2$\sigma$  criterion for PLCG identification. So, this galaxy could still have a buried AGN. The other two galaxies are at redshifts $z_{\rm spec}=2.442$ and 2.577. For these, the stellar bump centred at rest-frame $\sim 1.6 \, \rm \mu m$  is within the observed $5.8 \, \rm \mu m$ filter, producing a red $S_{5.8 \, \rm \mu m} / S_{3.6 \, \rm \mu m}$ colour. These two galaxies are close to the $S_{8.0 \, \rm \mu m} / S_{4.5 \, \rm \mu m}$ lower boundary of the Donley et al. colour wedge, so their presence within the wedge may simply be due to photometric scattering (note the size of the error bars).

Galaxy colour selections are attractive for their simplicity, but no criterion has been found to safely identify the bulk of the AGN population at any redshift. The PLCG criterion proposed here offers a suitable alternative, which is particularly successful at identifying AGN of different luminosities at $1.0<z<1.5$. Note that it would not be enough to extend the Donley et al.~(2012) to bluer colours, because most of the PLCGs display similar IRAC colours as other IR sources which do not have a significant power-law component in their SEDs, so an extended colour selection would produce a highly contaminated AGN sample.

\section{Summary and Conclusions}
\label{sec-summ}

In this paper I have studied the rest-frame  UV/optical/near-IR  ($\lambda_{\rm rest} \lsim 3 \, \rm \mu m$) SEDs of 174 $\tfm$-selected galaxies with secure spectroscopic redshifts $z_{\rm spec}>1$, analysing explicitly the presence and significance of a hot-dust, power-law component. I modelled the residual light produced after subtracting different possible power-law components, using stellar templates, on 11 broad-bands from the $B$-band through  $8 \, \rm \mu m$. The subtracted power-law components are characterised by three different spectral indices $\alpha=1.3$, 2.0 and 3.0, and different normalisation weights. 

I found that in 60 out of 174 (35\%) cases, the SED fitting with stellar templates without any previous power-law subtraction can be rejected with $> 2 \sigma$ confidence, i.e. 35\% of the $\tfm$ galaxies at $z>1$ are characterised by a significant power-law component in their rest near-IR SED. I referred to these galaxies as PLCGs. This high percentage  of PLCGs at $z>1$ appears to be inherent to the IR bright galaxy population. A similar analysis performed on a control sample of 247 IRAC-selected galaxies with $z_{\rm spec}>1$ and  similar stellar masses as the $\tfm$ galaxies, but which are not $\tfm$-detected, shows that the incidence of PLCGs is of only 18\%.

\vspace{0.2cm}

The incidence of PLCGs among $\tfm$ galaxies is much higher at $1.0<z<1.5$  (47\%) than at $z>1.5$ (21\%). The lower incidence at $z>1.5$ is explained by the fact that the presence of very hot dust ($T_{\rm dust} \gsim 1300 \, \rm K$) is necessary to have a significant power-law component manifested at observed wavelengths $\lambda \leq 8 \, \rm \mu m$. Therefore, only sources with the most extreme dust conditions would be recognised with the PLCG criterion at these higher redshifts.  

The impact of a significant power-law component in the rest near-IR SED on the derived galaxy stellar masses needs to be analysed on a case-by-case basis, but it is small in general. The correction for stellar masses is within a factor of 1.5 in $>80\%$ of cases. Only in a small percentage ($<10\%$) of the PLCGs, basically all corresponding to powerful AGN,  the correction to the stellar mass is so large that a stellar mass calculation based on the original photometry should clearly be avoided.

A key issue is understanding whether the PLCGs are truly related to the presence of an AGN, particularly at $1.0<z<1.5$, where the PLCG incidence is very high. To investigate this, I studied different properties of these sources. Firstly, I looked for identifications of the PLCGs in the ultra-deep 4Ms X-ray catalogue for the CDFS. I found that 77\% of the X-ray AGN  within my $\tfm$ sample with $1.0<z_{\rm spec}<1.5$ are classified as PLCGs. This indicates that the PLCG criterion provides a high completeness in selecting AGN at these redshifts. Interestingly, 78\% of the X-ray normal galaxies within my $\tfm$ sample with $1.0<z_{\rm spec}<1.5$ are also PLCGs. But their PLCG classification or X-ray luminosities do not correlate with the $\tfm$ luminosities,  so the X ray emission or the power-law SED excess is not simply a tracer of high star formation rates. 

A stacking analysis of the PLCGs that are not individually X-ray detected at $1.0<z<1.5$ reveals that they have on average similar properties as X-ray-detected normal galaxies. The stacked X-ray images yield a significant detection in the soft X-ray band ($0.5-2.0 \, \rm keV$), and no signal in the hard X-ray band ($2.0-8.0 \, \rm keV$). Around two thirds of the stacked PLCGs have a best-fitting power law with $\alpha=3$.

The visual inspection of optical  {\em HST}/ACS images has proven very useful to shed light on the nature of the PLCGs at $1.0<z<1.5$. I found a clear trend relating the PLCG morphology and the spectral-index $\alpha$ of the best-fitting SED power-law component: PLCGs with $\alpha=1.3$ or 2.0 look regular and typically have a nuclear matter concentration. Instead, the PLCGs with $\alpha=3.0$ typically show irregular morphologies, indicating a disturbed galaxy dynamics, which in some cases suggest the presence of galaxy mergers. This morphology trend is independent of the PLCG X-ray detection and classification. The $\alpha=3.0$ power-law component appears then to be a signature of quite extreme ISM conditions, where the interstellar dust is heated to temperatures $T \gsim 700 \, \rm K$ by the gas shocks produced in the disturbed ISM.

Of course, this does not preclude the  $\alpha=3.0$ PLCGs to also host AGN. Actually, the presence of an AGN is confirmed in some of them by the X-ray data. But an AGN presence does not appear to be necessary to produce a significant $\alpha=3.0$ power-law component.  For the PLCGs with $\alpha=1.3$ or 2.0, instead, higher dust temperatures are necessary $T \gsim 900 \, \rm K$. This property, along with the regular morphologies, take to the direct conclusion that these sources must host an AGN. One could speculate that the $\alpha=3.0$ PLCGs may all be on the way to form an AGN as well, but this hypothesis cannot be tested with the current data.

Considering that all the PLCGs with $\alpha=1.3$ or 2.0 in my sample contain AGN, and adding all the other X-ray classified AGN within my $\tfm$ sample, I obtain that a total of 30\% of the $\tfm$ sources at $1.0<z<1.5$ host AGN. This should be considered as a lower limit, as some of the $\alpha=3.0$ PLCGs that are not individually X-ray detected could also contain AGN. If all of them had AGN, that would give an upper limit of $\sim 52\%$ on the AGN fraction among LIRGs and ULIRGs at $1.0<z<1.5$. Very recently, different works have concluded on a higher fraction of AGN among IR galaxies than previously known at $z<1$, obtaining percentages of 37-50\% \citep[e.g.][]{alo12,jun13}. The results of my study are in line with such conclusions. At $z>1.5$, a similar criterion as that used at $z<1.5$ would indicate that $\sim 40\%$ of the $\tfm$ galaxies contain an AGN, although this figure is largely based on the X-ray detections, rather than the PLCG incidence, which is relatively low (for a comparison, Pozzi et al.~2012 derived a fraction of 35\% based on the mid-/far-IR SED analysis of 24 ULIRGs at $z\sim2$).

In this work, I have calibrated the PLCG analysis technique making use of galaxies with secure spectroscopic redshifts. A complete analysis of the applicability of this technique with the simultaneous determination of photometric redshifts ($z_{\rm phot}$) will be presented in a future paper. A preliminary run of my SED fitting code leaving the redshift as a free parameter indicates that, in fact, the consideration of a power-law subtraction produces some improvement in the overall $z_{\rm phot}$-$z_{\rm spec}$ comparison, reducing the number of catastrophic outliers. This suggests that the identification of PLCGs can be done, and the power-law subtraction is actually recommendable, when attempting a $z_{\rm phot}$ determination.

Although less straightforward than a simple IRAC colour-colour selection, the PLCG identification criterion introduced here is much more complete to select AGN of different luminosities, especially in the redshift range $1.0<z<1.5$. Keeping only those PLCGs with spectral index $\alpha=1.3$ and 2 should result in the best compromise between completeness and reliability, as most X-ray normal galaxies and X-ray non-detections are among the PLCGs with $\alpha=3$. In the absence of very deep X-ray data, the PLCG selection with spectral index segregation offers the most efficient method to identify the AGN presence in any large galaxy sample.

\acknowledgments

Based on observations made with the {\em Spitzer Space Telescope}, which is operated by the Jet Propulsion Laboratory, California Institute of Technology under a contract with NASA. Also based on observations undertaken at the European Southern Observatory (ESO) {\em Very Large Telescope (VLT)} under different programmes; the NASA/ESA {\em Hubble Space Telescope}, obtained at the Space Telescope Science Institute, operated by the Association of Universities for Research in Astronomy, Inc. (AURA), under NASA contract NAS 5-26555; and the {\em Chandra X-ray Observatory}, which is operated by the Smithsonian Astrophysical Observatory for and on behalf of NASA under contract NAS8-03060.

I thank Almudena Alonso-Herrero, Maurilio Pannella, Brigitte Rocca-Volmerange, and John Silverman for very useful discussions, and an anonymous referee for a very constructive report. I am also grateful to the GOODS Team for the public release of their superb data products.

{\it Facilities:} \facility{{\em Spitzer} (IRAC,MIPS)}, \facility{{\em HST} (ACS)}, \facility{VLT (ISAAC, FORS2, VIMOS)}.

\end{document}